\documentclass[aps,prx,twocolumn,longbibliography,superscriptaddress]{revtex4-1}

\usepackage{dcolumn}
\usepackage{bm}
\usepackage[english]{babel}
\usepackage{lipsum} 
\usepackage{times}
\usepackage{amsfonts,amssymb,bm}
\usepackage{epsfig}
\usepackage{mathrsfs}
\usepackage{color,soul}
\usepackage[normalem]{ulem}
\usepackage{bbold}
\usepackage{pstricks}
\usepackage{multirow}
\usepackage{graphicx}
\usepackage{textcomp}
\usepackage{braket}
\usepackage{empheq}
\usepackage{hyperref}
\hypersetup{colorlinks,%
citecolor=red,%
filecolor=black,%
linkcolor=blue,%
urlcolor=blue}
\usepackage{color}
\usepackage{rotating}
\usepackage{amsthm}
\usepackage{siunitx}
\usepackage{amsmath}
\usepackage{eurosym}
\usepackage[babel]{csquotes}

\usepackage[font=small]{caption}
\usepackage{mwe}

\usepackage[framemethod=default]{mdframed}
\usepackage[labelfont=bf,labelsep=quad,justification=raggedright]{caption}
\usepackage{newfloat}

\DeclareFloatingEnvironment[name={S}]{suppfigure}
\DeclareFloatingEnvironment[name={S}]{supptable}
\allowdisplaybreaks

\definecolor{green1}{rgb}{0.33, 0.7, 0.69}

\newmdenv[linecolor=white,backgroundcolor=gray!15!]{myframe}

\AtBeginDocument{%
    \newwrite\bibnotes
    \def\bibnotesext{Notes.bib}
    \immediate\openout\bibnotes=\jobname\bibnotesext
    \immediate\write\bibnotes{@CONTROL{REVTEX41Control}}
    \immediate\write\bibnotes{@CONTROL{%
    apsrev41Control,author="08",editor="1",pages="1",title="0",year="1"}}
     \if@filesw
     \immediate\write\@auxout{\string\citation{apsrev41Control}}%
    \fi
}%
        
\begin{document}
\title{Qubits are not observers -- a no-go theorem}

\author{\v{C}aslav Brukner}
\affiliation{Vienna Center for Quantum Science and Technology (VCQ), Faculty of Physics, University of Vienna, Boltzmanngasse 5, 1090, Vienna, Austria}
\affiliation{Institute for Quantum Optics and Quantum Information (IQOQI-Vienna), Austrian Academy of Sciences, Boltzmanngasse 3, 1090 Vienna, Austria}



\begin{abstract}

The relational approach to quantum states asserts that the physical description of quantum systems is always relative to something or someone.   In relational quantum mechanics (RQM) it is relative to other quantum systems, in the (neo-)Copenhagen interpretation of quantum theory to measurement contexts, and in QBism to the beliefs of the agents.  In contrast to the other two interpretations, in RQM any interaction between two quantum systems counts as a "measurement", and the terms "observer" and "observed system" apply to arbitrary systems.  We show, in the form of a no-go theorem, that in RQM the physical description of a system relative to an observer cannot represent knowledge about the observer in the conventional sense of this term. The problem lies in the ambiguity in the choice of the basis with respect to which the relative states are to be defined in RQM. In interpretations of quantum theory where observations play a fundamental role, the problem does not arise because the experimental context defines a preferred basis. 

\end{abstract}

\maketitle

The method of thought experiment allows to consider hypothetical situations in order to analyze the internal consistency of the tested theories. In quantum theory an interesting situation arises when the observer is considered to have quantum properties. This is not problematic in all approaches where quantum theory is assumed to in principle describe {\em anything}. (Although according to some interpretations it cannot encompass {\em everything}, to put it in Peres' terms~\cite{peres}). The explicit quantum description of an observer is at the heart of the Wigner’s friend thought experiment~\cite{wigner}. Recent theoretical results~\cite{brukner, frauchinger,brukner1,bub, bub1, healey, bong,baumann,debrota,allard,eric} and experimental tests~\cite{proletti,bong} on the extended version of the thought experiment restrict the possibility of maintaining the view that quantum states (and with it the observed events) are absolute and not relative to anything or anyone.

Relational quantum mechanics (RQM) ~\cite{rovelli,enc,biagio}, the neo-Copenhagen interpretation~\cite{bub,bub1,healey,brukner}, and QBism~\cite{qbism} all share the view that quantum state of a system is defined only relationally. In RQM it is relative to another system, in the neo-Copenhagen interpretation to the experimental context, and in QBism to the personal beliefs of the agent. Unlike the other two interpretations, however, RQM makes no reference to the concepts of "measurement," "observation," or "agent" when analyzing the relational aspect of quantum states. All that is necessary in RQM for a quantum system to have defined properties relative to another system is the existence of an interaction between the two systems. A physical interaction is the establishment of a correlation between quantum systems, where one system is colloquially referred to as the "observer" and the other as the "observed system." As an example, in RQM it is meaningful to speak of quantum state of an electron relative to another in a helium atom.

In this short note, I will derive a no-go theorem that restricts the possibility of understanding the relational description in RQM as knowledge that one system can have about another in the conventional sense of that term. The key observation is that two quantum systems that are in a joint state can have different correlations depending on the basis chosen to represent the state. This directly implies that in RQM an "observer" can have multiple relative state assignments to the same "system". Moreover, different states of the observer's knowledge are not orthogonal to each other and thus cannot be copied or used to condition the agent's actions, which raises the question in what sense they can be associated with the notion of "knowledge". The problems do not arise when the "observer" is identified by a (macroscopic) measurement device or memory of an agent, and only the correlations in a single, preferred basis account for the definition of the relative quantum state. Since this is the case in the neo-Copenhagen and also compatible with the QBist interpretation of quantum theory, our results can be summarized as drawing a demarcation line between RQM and the other two interpretations advocating relative quantum states. 

To illustrate a typical situation where a relational description of a physical system is apparent, we consider two quantum systems, one of which we will refer to as the "system" (S) and the other as the "observer" (O). Specifically, we consider the following entangled state between the observer and the spin-1/2 particle: \begin{equation}
    |\psi\rangle_{SO}= \frac{1}{\sqrt{2}} (|\uparrow\rangle_S |\uparrow\rangle_O +|\downarrow\rangle_S |\downarrow\rangle_O), \label{meas}
\end{equation}
where $|\!\uparrow\rangle_S$ and $|\!\downarrow\rangle_S$ are the system's states "spin up" and "spin down" and $|\uparrow\rangle_O$ and $|\downarrow\rangle_O$ are the corresponding states of the observer, which are correlated to the "spin up" and "spin down" states, respectively.
If the observer is in state $|\!\uparrow\rangle_O$, she "knows" that the system is in state $|\!\uparrow\rangle_S$, and if she is in state $|\!\downarrow\rangle_O$, she " knows" that the system is in state $|\!\downarrow\rangle_S$. Either way, the system is in a definite state of spin-up/down relative to the observer. When the correlations in Eq.~(\ref{meas}) are established in a measurement, the observer's states can be identified with definite pointer states of a measuring device or definite perceptual states of the observer.

Note that state~(\ref{meas}) can also be written in a different basis associated to S and O as
\begin{equation}
  |\psi\rangle_{SO}= \frac{1}{\sqrt{2}} (|\rightarrow\rangle_S |\rightarrow\rangle_O +|\leftarrow\rangle_S |\leftarrow\rangle_O), \label{other}
\end{equation}
where $|\rightarrow\rangle_i = \frac{1}{\sqrt{2}} (|\uparrow\rangle_i + |\downarrow\rangle_i)$, $|\leftarrow\rangle_i = \frac{1}{\sqrt{2}} (|\uparrow\rangle_i - |\downarrow\rangle_i)$ with $i=S, O$. Here $|\!\rightarrow\rangle_S$ and $|\!\leftarrow\rangle_S$ are states of "spin right" and "spin left" of the system, respectively.   

The two possible basis representations of state $|\psi\rangle_{SO}$ give rise to the so-called preferred basis problem. What would be the state of  system S relative to O: the one with a definite spin up/down as in Eq.~(\ref{meas}), or the one with a definite spin right/left as in Eq.~(\ref{other})? (For the state there are actually infinitely many such basis decompositions). RQM does not seem to give any prescription on how to resolve this ambiguity (See Refs.~\cite{mucino,rovelli1} for a recent discussion of the issue). 

However, as assumed in RQM, we can use {\em every} correlation (of type (\ref{meas}) and (\ref{other})) to define what O "knows" about S. As we will see below, this becomes inconsistent under a set of reasonable assumptions defining the nature of such "knowledge". The proof is a simplified version of that of Bassi and Ghirardi from Ref.~\cite{bassi}. 

When formalizing our no-go theorem it is sufficient to consider finite dimensional Hilbert spaces for our purposes. We start with stating the assumptions of the no-go theorem.

\begin{description}

\item[\textsc{1. Definite Relative State (DefRS)}] For {\em any} set of states $\{|x_i\rangle_S, |X_i\rangle_O\}$, $i=1,2,...$, of the system S and the observer O such that 
\begin{equation}
    |\psi\rangle_{SO} = \sum_i c_i |x_i\rangle_S |X_i\rangle_O, \label{axiom}
    \end{equation}
($c_i\in \mathbb{C}$ and $\sum_i |c_i|^2=1$) holds for a given joint state $|\psi\rangle_{SO}$, the states $|X_i\rangle_O$ are states of knowledge of the observer. When the observer is in state $|X_i\rangle_O$, she knows that the system is in the definite relative state $|x_i\rangle_S$. 

\item[\textsc{2. Distinct relative States (DisRS)}]
The observer's states of knowledge $|X\rangle_O$ and $|X'\rangle_O$, which are correlated with distinct relative states $|x\rangle_S$ and $|x'\rangle_S$ of the system, are represented by orthogonal vectors in the observer's Hilbert space, i.e. if $|x\rangle_S \neq |x'\rangle_S$, then $\langle X|X'\rangle = 0.$
\end{description} 
Note that in \textsc{DisRS} the states of the system need not be orthogonal, but only different from each other.

We give a justification for the formulation of the two assumptions. The assumption \textsc{DefRS} formalizes the commonly held view that any information one system has about another exists because of correlations between them. 
The justification of the \textsc{DisRS} assumptions is best seen when we consider correlations between the states of the system and the observer to be established in a measurement. One can then identify the states of the observer's memory that are correlated to different system's states, as states of {\em distinguishable perceptions}. Distinguishable perceptions are experienced by observers as distinguishable sensations and, they can also be identified as distinguishable brain activities from outside the observers. In quantum theory, distinguishable states are represented by orthogonal vectors. 

We will show that the two assumptions cannot be both fulfilled. We start by assuming that that the observer assigns a definite relative state to a spin-1/2 particle in one specific basis in agreement with \textsc{DefRS} assumption. To put it concretely, we assume that the joint state of the observer and the spin particle is: 
\begin{equation}
     |\psi\rangle_{SO}= \frac{1}{\sqrt{2}} (|\uparrow\rangle_S |U\rangle_O +|\downarrow\rangle_S |D\rangle_O), \label{srce}
\end{equation}
where we introduce  states $|U\rangle_O$ and $|D\rangle_O$ to indicate observer's "knowledge" or "perception" of "spin up" and "spin down" respectively. Due to \textsc{DisRS} assumption one has $\langle U|D\rangle =0$.

Note that state~(\ref{srce}) can be rewritten as
\begin{eqnarray}
  |\psi\rangle_{SO}  &=& \frac{1}{2} [|\rightarrow\rangle_S (|U\rangle_O + |D\rangle_O) \nonumber \\ & & +|\leftarrow\rangle_S (|U\rangle_O - |D\rangle_O)]. \label{cat}
\end{eqnarray}
in the basis introduced below Eq.~(\ref{other}). Since this expression is also of the form~(\ref{axiom}), we again apply \textsc{DefRS} and conclude that in the states $|\pm \rangle_O \equiv \frac{1}{\sqrt{2}}(|U\rangle_O \pm |D\rangle_O)$ the observer knows the relative state "spin left/right" of the system. This relative state is different from "spin up/down". Thus, according to the \textsc{DisRS} assumption, the corresponding states $|\pm \rangle_O$ of observer's knowledge should be represented by vectors orthogonal to $|U\rangle_O$ and $|D\rangle_O$. However, an explicit calculation shows that $|\langle \pm |U\rangle| = |\langle \pm| D\rangle| =\frac{1}{\sqrt{2}}$. (Note that for the same reason, the two states of knowledge are also not identical to $|U\rangle_O$ or $|D\rangle_O$.) This concludes the proof.

We arrive at the conclusion that at least one of the assumptions cannot be valid in quantum mechanics. We now analyze what are the consequences of violating each of the two assumptions. 

We start with the assumption \textsc{DefRS}. Its violation is consistent with the view that the states of knowledge of the observer are defined in terms of correlations between the system and the observer in only one, {\em preferred}, basis. Only with respect to this one specific basis can the observer assign a relative state to the system. More precisely, a measurement device or observer's memory have many degrees of freedom whose states, we assume, can be divided into equivalence classes, each class being specified by a particular perception. Most of these degrees of freedom are microscopic degrees of freedom over which one has no or very limited control. Accordingly, we shall indicate the state of observer's memory as $|X, \alpha\rangle$ where $X$ is a label which indicates the specific perception and $\alpha$ denotes all its uncontrollable and unknown degrees of freedom. For example, $X$ can be $U$ to indicate perception of "spin up", $D$ for "spin down" etc. In agreement with \textsc{DisRS} one has $|\langle U,\alpha|D,\alpha\rangle| \approx 0.$ (Note that one may have $\langle X,\alpha|X,\alpha'\rangle$ = 0 since orthogonal states can be assigned to the same perception if microscopic degrees of freedom are in orthogonal states.) In agreement with the no-go theorem the states $\{|U,\alpha\rangle_O,|D,\alpha\rangle_O\}$ define the preferred basis, while states $\{|\pm,\alpha\rangle_O \equiv \frac{1}{\sqrt{2}}(|U,\alpha\rangle_O \pm |D,\alpha\rangle_O) \}$ do not represent any state of a definite perception. The existence of the preferred basis is also supported by the research in decoherence theory~\cite{rmp}. It is interesting to note, however, that already reasonable assumptions about the nature of the knowledge an observer should have about what is observed presuppose the existence of a preferred basis. Finally, we note that the neo-Copenhagen interpretation of quantum theory is consistent with a violation of \textsc{DefRS}, since there a quantum state is always given relative to the context of the measurement, which in turn defines the preferred basis. The QBist view may also be (made) compatible with a violation of \textsc{DefRS} if a purified description of the measurement is admissible within this view (at least as a valid description of a second, external agent) and quantum states are defined relative to the agent's perceptions.

Violation of the assumption \textsc{DisRS} would imply a departure from the idea that distinct quantum states of the observer's knowledge should correspond to distinguishable perceptions and therefore be represented by mutually orthogonal vectors. However, one might imagine that the kind of "knowledge" that one quantum system can have about another, based solely on correlations between them, is different from what we normally think of as knowledge. For example, the Heisenberg uncertainty principle limits the amount of knowledge an agent can have about complementary observables, such as position and momentum. Here the agent's knowledge is conditioned by the information stored in classical memory. In contrast, a "quantum agent", simply by virtue of its correlations with another quantum system, may somehow simultaneously "know" the values of mutually complementary observables~\cite{renato}. Obviously, such "quantum knowledge" would violate the Heisenberg uncertainty principle. In fact, one could view the "quantum agent" as a quantum memory entangled with the system, which could be "accessed" to accurately predict the values of complementary observables simultaneously~\cite{renner2}. It remains open, of course, what it would mean operationally to "access" such knowledge or to "make prediction" on the basis of it. We conclude that RQM might preserve the ontology of relations between quantum systems but only at the expanse of a radical revision of the notion of "knowledge about a physical system". Finally, we note that both assumptions might be invalid.

The no-go theorem implies that a superposition state of definite perceptions is not a state of a definite perception. But if so, what is it? In other words, {\em what is it like to be a Schrödinger cat?} to paraphrase and extend the famous question {\em "What is it like to be a bat?"} from the title of a paper by Nagel~\cite{nagel} into the quantum domain (See~\cite{lewis,nikolic} for two papers in which the same question has already been asked and an answer offered.) Note that such a superpositon state can in principle be prepared by projecting the spin system in the left/right basis in the state~(\ref{cat}). One possible answer to the above question, which we endorse here, is that the observer in the superposition state would nonetheless perceive a well-defined spin up/down. This is  consistent with the observer still being in the superposition state as described by another, external observer (Wigner) who assigns the state to a well-defined experimental procedure for its preparation and verification. 

In the conclusions, our results show that not every quantum system can serve as a measuring device or observer. This could be seen as contradicting one of the main premises of relational quantum mechanics about the equivalence of all quantum systems. While every measurement can be seen as an interaction between two systems (e.g., as a valid description of the measurement of the friend from Wigner's point of view), not every interaction is a measurement. It becomes one when the observer's states, that are correlated with different states of the system, are mutually distinguishable, can be copied and unambiguously communicated, like any knowledge. Qubits are not observers.


{\em Note added}: after the completion of this work I became aware of an independent study on the same topic by Jacques L. Pienaar, see arXiv:2107.00670.

I thank Philippe Allard Guerin, Veronika Baumann, Flavio Del Santo and Martin Renner for interesting discussions. I acknowledge  financial support from the Austrian Science Fund (FWF) through BeyondC (F7103-N48) and project no. I-2906, from the European Commission via Testing the Large-Scale Limit of Quantum Mechanics (TEQ) (No.  766900) project, and from Foundational Questions Institute (FQXi).  This research was supported by the John Templeton Foundation grant (ID 61466) as part of the The Quantum Information Structure of Spacetime (QISS) Project (qiss.fr).


\begin{thebibliography}{100}

\bibitem{peres} A. Peres, Quantum Theory: Concepts and Methods (Springer: New York, NY, USA, 1995; p. 173).

\bibitem{wigner} E. P. Wigner, Remarks on the mind-body question, 
in The Scientist Speculates, Ed. I. J. Good (Heinemann: London, UK, 1961).

\bibitem{brukner} {\v C}. Brukner, On the quantum measurement problem, 
in Quantum [Un]speakables II; Eds. R. Bertlmann, A. Zeilinger (The Frontiers Collection; Springer: New York, NY, USA, 2016).

\bibitem{frauchinger} D. Frauchiger and R. Renner, Quantum theory cannot consistently describe the use of itself, Nat. Commun. {\bf 9}, 3711 (2018).

\bibitem{brukner1} {\v C}. Brukner, A no-go theorem for observer-independent facts, Entropy {\bf 20}, 350 (2018).

\bibitem{healey} R. Healey, The Quantum Revolution in Philosophy (Oxford: Oxford University Press, 2017).

\bibitem{bub} J. Bub, Why Bohr was (Mostly) Right, arXiv:1711.01604v1 (2017).

\bibitem{bub1} J. Bub,  In defense of a ``single-world'' interpretation of quantum mechanics, Stud. Hist. Philos. Mod. Phys. {\bf 72},
251–255 (2020). 

\bibitem{bong} K. W. Bong, A. Utreras-Alarcon, F. Ghafari, Y. C. Liang, N. Tischler, E. G. Cavalcanti, G. J. Pryde and H. M. Wiseman, A strong no-go theorem on the Wigner’s friend paradox, Nat. Phys. {\bf 16}, 1199–1205 (2020).

\bibitem{baumann} V. Baumann and {\v C}. Brukner, Wigner's friend as a rational agent, in: Hemmo M., Shenker O. (eds) Quantum, Probability, Logic, Jerusalem Studies in Philosophy and History of Science (Springer, Cham, 2020).

\bibitem{debrota} J. B. DeBrota, C. A. Fuchs and R. Schack, Respecting One's Fellow: QBism's Analysis of Wigner's Friend, arXiv:2008.03572 (2020).

\bibitem{allard} P. Allard Guerin, V. Baumann, F. Del Santo and {\v C}. Brukner, A no-go theorem for the persistent reality of Wigner’s friend’s perception, Commu. Phys. {\bf 4}, 93 (2021).

\bibitem{eric} E. Calvalcanti, The view from a Wigner bubble, Found. Phys. {\bf 51}, 39 (2021).

\bibitem{proletti} M. Proietti, A. Pickston, F. Graffitti, P. Barrow, D. Kundys, C. Branciard, M. Ringbauer and A. Fedrizzi, Experimental test of local observer independence, Sci. Adv. {\bf 5}, eaaw9832 (2019).

\bibitem{rovelli} C. Rovelli, Relational Quantum Mechanics, Int. J. of Theor. Phys., {\bf 35} 1637-1678 (1996).

\bibitem{enc} F. Laudisa and C. Rovelli, Relational Quantum Mechanics, The Stanford Encyclopedia of Philosophy, Edward N. Zalta (ed.) (2019 Edition).

\bibitem{biagio} A. Di Biagio and C. Rovelli, Stable Facts, Relative Facts, Found. Phys. {\bf 51}, 30 (2021).

\bibitem{qbism} C. A. Fuchs, Notwithstanding Bohr, the Reasons for QBism, Mind Matter {\bf 15}, 245–300 (2017).

\bibitem{mucino} R. Mucino, E. Okon and D. Sudarsky, Assessing Relational Quantum Mechanics, arXiv:2105.13338 (2021).

\bibitem{rovelli1} C. Rovelli, A response to the Mucino-Okon-Sudarsky’s Assessment of Relational Quantum Mechanics, arXiv:2106.03205 (2021)

\bibitem{rmp} M. Schlosshauer, Decoherence, the measurement problem, and interpretations of quantum mechanics, Rev. Mod. Phys. {\bf 76}, 1267 (2005).

\bibitem{renato} R. Renner, privite communication.

\bibitem{renner2} M. Berta, M.  Christandl, R. Colbeck, J. M. Renes and R. Renner, The uncertainty principle in the presence of quantum memory, Nat. Phys. {\bf 6}, 659–662 (2010).

\bibitem{bassi} A. Bassi and G. Ghirardi, A General Argument Against the Universal Validity of the Superposition Principle, Physics Letters A {\bf 275}, 373--381 (2000). 

\bibitem{nagel} T. Nagel, What Is It Like to Be a Bat?, The Philosophical Review  {\bf 83} (4), 435--450 (1974).

\bibitem{lewis}  P. J. Lewis, What is it like to be Schrödinger's cat?, Analysis {\bf 60}, No. 1, 22--29 (2000).

\bibitem{nikolic} H. Nikolic, Internal environment: What is it like to be a Schrodinger cat?, Eur. J. Phys. {\bf 36}, 045003 (2015).

\end{thebibliography}
\end{document}